\documentclass[aps,prd,amssymb,showpacs,floatfix,twocolumn,superscriptaddress,nofootinbib]{revtex4-1}
\usepackage{amsmath,amsfonts,graphics,epsfig,amssymb}

\begin{document}
\title{Towards discrimination between galactic and intergalactic
axion--photon mixing}

\author{Sergey Troitsky}

\email{st@ms2.inr.ac.ru}

\affiliation{Institute for Nuclear
Research of the Russian Academy of Sciences, 60th October Anniversary
Prospect 7a, Moscow 117312, Russia}



\begin{center}
\begin{abstract}
There exists a growing evidence for the
anomalous transparency of the Universe for energetic gamma rays. Popular
explanations include conversion of photons into hypothetical axion-like
particles (ALPs) and back in astrophysical magnetic fields. Two
distinctive scenarios of this conversion have been put forward: either it
happens in the (host galaxy of the) gamma-ray source and in the Milky Way,
or the photon-ALP oscillations take place in the intergalactic magnetic
fields all along the way between the source and the observer. Here we
point out that, given recent astrophysical constraints on ALPs and on
intergalactic magnetic fields, these two mechanisms imply very different
ALP masses and couplings. Therefore, confirmation of the anomalies and
identification of one of the scenarios would mean cornering of ALP
parameters to a particular narrow region. We discuss approaches to
distinguish between the two mechanisms and present some preliminary
indications in favour of the galactic scenario.
\end{abstract}
\end{center}
\maketitle

\section{Introduction}
\label{sec:intro}
Since the early years of very-high-energy (VHE, above 100~GeV) gamma-ray
astronomy, more and more distant sources have been discovered in this
energy band, see e.g.\ Refs.~\cite{3c279, hardening, 2photons}. This was
surprising because gamma rays of these energies should interact with the
background radiation to produce electron-positron pairs \cite{Nikishov},
the process which results in a strong suppression of the flux from distant
sources. The apparent weakness of the suppression might be attributed
either to overestimation of the amount of background radiation or to
peculiar emission or absorption mechanisms at work in particular sources.
However, recent studies indicate
that these explanations are hardly relevant: modern models of
extragalactic background light, e.g.\ Refs.~\cite{Franceschini, Gilmore},
used in the studies, already saturate lower bounds~\cite{EBLlower-bounds}
from simple galaxy counts; while analyses of
sets of distant sources~\cite{HornsMeyer, gamma, Galanti-spindex} revealed
unphysical redshift dependence in the required peculiar features of the
emission spectra.

One of the most intriguing explanations of the ensemble of the data is
that we might observe effects of a new hypothetical axion-like particle
(ALP) (see Ref.~\cite{Ringwald-review} for a review). Depending on
parameters of the ALP and on values of the intergalactic magnetic field
(IGMF), two scenarios may work. Photon-ALP oscillations in IGMF may
effectively increase the photon mean free path \cite{Csaba, DARMA,
DARMA1}.
Alternatively, if IGMF is weak, photons may efficiently convert to ALPs in
the source \cite{Serpico} or in its neighbourhood, cluster or filament
\cite{FRT}, with reconversion back to photons in, or close to, the Milky
Way. The main purpose of this study is to emphasise differences between
the two scenarios and to demonstrate that, given recent astrophysical
constraints, they correspond to distinguishable ALP
parameters. The key new ingredients contributing here are the reanalysis
of the supernova (SN) 1987A data~\cite{SN1987Anew} and new upper limits on
intergalactic magnetic fields~\cite{PshirkovIGMF}. We also present some
evidence in favour of the galactic scenario and discuss future
observations which may help to discriminate between the two options (see
also Refs.~\cite{Meyer-evidence, HESSirreg}).

The rest of the paper is organized as follows. In Sec.~\ref{sec:evidence},
we briefly review the evidence for the anomalous transparency of the
Universe for gamma rays. In Sec.~\ref{sec:explain}, we describe two
scenarios of the ALP-photon mixing which may explain the observed
evidence. Then, we compare required ALP parameters, calculated by means of
detailed analyzes elsewhere, with recent astrophysical bounds and
present argumentation for the main point of the paper. In
Sec.~\ref{sec:discri}, we consider ways to distinguish the two scenarios
and present some evidence in favour of one of them. Future prospects to
test the ALP explanation of the anomalous transparency of the Universe for
gamma rays and to single out one of the scenarios, thus constraining the
ALP parameters, are discussed in Sec.~\ref{sec:future}.

\section{Evidence for anomalous transparency}
\label{sec:evidence}
The modern evidence for the anomalous transparency of the Universe for
energetic gamma rays is based on studies of ensembles of distant VHE
sources. The observed spectra of these sources have been corrected for the
pair-production effects (``deabsorbed'') within the most conservative,
i.e.\ lowest-absorption, models, to obtain intrinsic spectra emitted
at the sources. These intrinsic spectra exhibit unphysical redshift
dependence which is readily interpreted as an overestimation of the
absorption even in the minimal models.

Upward breaks, or unusual spectral hardenings, have been found in
{\it deabsorbed} spectra of many individual sources (see e.g.\
Ref.~\cite{hardening}). Statistically significant hardening never presents
in observed spectra which, contrary, often exhibit mild softening at
high energies. In 2012, Horns and Meyer analysed \cite{HornsMeyer} a
sample
of 7 blazars observed at optical depths $\tau>2$ with respect to the pair
production. The blazars were observed by imaging atmospheric Cerenkov
telescopes (IACTs) and have redshifts $z\lesssim 0.536$. They found an
evidence that positions of the upward breaks in deabsorbed spectra of
blazars change with redshift in such a way that they always occur at the
energy where the absorption becomes important. This was surprising because
astrophysical properties of blazars in the sample were very similar for
close and distant sources. The probability that this effect is a chance
fluctuation, estimated in Ref.~\cite{HornsMeyer} by a statistical
procedure based on the Kolmogorov-Smirnov test, corresponds to that of a
$4.2\sigma$ fluctuation in a Gaussian distribution.

Rubtsov and Troitsky \cite{gamma} considered in 2014 a sample of 20
blazars observed at optical depths $\tau>1$. IACT results have been
supplemented with the FERMI-LAT data, which allowed to extend the redshift
range up to $z\approx 2.156$. Assuming the breaks in deabsorbed spectra at
energies for which $\tau=1$, we found that the strength of the break, that
is the difference between the power-law spectral indices below and above
the break point, is a function of the redshift and does not depend on the
physical properties of a blazar. The statistical significance of this
unphysical dependence, indicating overestimation of the absorption and
therefore an anomaly, has been calculated in Ref.~\cite{gamma} by means of
the usual $\chi^{2}$ analysis and corresponds to a $12.4\sigma$
fluctuation of a Gaussian.

In 2015, Galanti et al.\ \cite{Galanti-spindex} considered a sample of 39
blazars ($z \lesssim 0.536$)
detected in VHE gamma rays, independently of the opacities
tested. They described deabsorbed spectra as power-law functions and did
not fit the spectral breaks. The power-law spectral index was found to be
redshift-dependent, which is not expected in any astrophysical model and
again indicates the anomaly. This result confirms the observation of
Ref.~\cite{gamma} though the intrinsic scatter of spectral indices and the
limited redshift range make the result less pronounced than in the break
study.

The significance estimates quoted above are based on statistical analyses
only and are subject to systematic uncertainties, which are discussed in
Refs.~\cite{HornsMeyer, gamma}. In particular, Ref.~\cite{HornsMeyer}
shows that under the worst assumptions about systematic errors, the
significance of the effect is reduced by $\sim 1.6 \sigma$.
Ref.~\cite{gamma} presents several tests of the robustness of the result;
however, a detailed quantitative study of systematic uncertainties is
hardly possible without a complete sample of sources (tests of the
Malmquist bias have been presented in Ref.~\cite{gamma} for the part of
the sample taken from the FERMI-LAT catalog).

Recently, Biteau and Williams~\cite{crit} criticised the results of Horns
and Meyer~\cite{HornsMeyer} and claimed they do not see any anomaly in the
ensemble of deabsorbed spectra of 38 blazars observed by IACTs. Since they
did not select the blazars by the energies at which the data are available,
their sample is dominated by the sources for which the absorption is, and
should be, low or negligible (as we know from Ref.~\cite{gamma}, only 15
blazars were firmly observed by IACTs at $\tau>1$, and~\cite{HornsMeyer}
only 7 at $\tau>2$). In addition, they used multiple spectra for a given
blazar in their work; this introduces additional statistical weight to
better-studied nearby, and therefore unabsorbed, sources. Even for heavily
absorbed sources, they apply a different method to derive their own
model of background radiation and the intrinsic spectrum. All these points
might explain the discrepancy between Refs.~\cite{crit} and
\cite{HornsMeyer} which Biteau and Williams attribute to the use of the
Kolmogorov-Smirnov test by Horns and Meyer. In any case, this critique is
irrelevant to our work~\cite{gamma} in which we made use of the usual
$\chi^{2}$ test to determine the significance of the anomalous
transparency in a clean sample.

The only astrophysical explanation of these anomalies~\cite{Kusenko}
requires some non-conventional assumptions. It assumes that a sufficient
number of ultra-high-energy cosmic protons are accelerated in precisely
the same gamma-ray blazars. Unless extragalactic magnetic fields are
as low as $\lesssim 10^{-17}$~G everywhere along the line of sight, this
scenario may have tensions with the observation of fast variability of
4C$+21.35$ at very high energies~\cite{PKS1222+21}. In what follows, we
assume the anomalies are real and concentrate on a different kind of their
explanations.

\section{Explanations}
\label{sec:explain}
The problem of unphysical spectral breaks may be alleviated if the
gamma-ray attenuation is reduced by means of some mechanism. Indeed, the
breaks are seen precisely at the energies where the correction for
attenuation becomes important (lower energies for larger distances);
reduction of the attenuation diminished the correction and removes the
breaks~\cite{gamma}: given observed softening of high-energy spectra, one
may even obtain expected power-law shapes if the absorption is present but
reduced. However, the usual deabsorption procedure is based on firm
standard physics and very conservative assumptions about the photon
background, so only new-physics effects might help. Besides the
possibility of the Lorentz-invariance violation, the only known
explanation involves ALPs.

An ALP mixes with photons in external magnetic
field~\cite{RaffeltStodolsky}, which may allow to suppress the attenuation
due to pair production: gamma-ray photons convert to ALPs, then travel
unattenuated and eventually convert back to photons. The photon beam is
still attenuated, but the flux suppression becomes less severe. A useful
collection of formulae describing the mixing for astrophysical
environments, as well as a quantitative discussion of various scenarios,
may be found in Ref.~\cite{FRT}. To reduce the opacity of the Universe for
TeV gamma rays from blazars, two particular scenarios involving ALPs are
important. The purpose of the present study is to emphasise and to explore
the difference between the two approaches.

The first scenario implies that the intergalactic magnetic field is strong
enough to provide for ALP/photon conversion all along the path between the
source and the observer. Originally suggested in Ref.~\cite{Csaba} in a
different context, this mechanism, known also as the DARMA scenario, was
invoked for the TeV blazar spectra in Ref.~\cite{DARMA}.
If it is at work, then the photon/ALP mixed beam propagates through the
Universe and, since the photons are attenuated while ALPs are not, the
effective suppression of the flux is smaller compared to the pure-photon
case.
It is easy to demonstrate that, for a sufficiently long
propagation through domains of randomly oriented magnetic fields, the
optical depth is effectively reduced by 2/3 in this scenario. A more
detailed study is given, for instance, in Ref~\cite{Galanti-spindex},
from where the most recent constraints on the relevant ALP parameters are
obtained: the ALP mass $m \lesssim
10^{-9}$~eV and the ALP-photon coupling $g_{a\gamma \gamma}$ determined
from $\xi \equiv (B/{\rm nG}) (g_{a\gamma \gamma} \times 10^{11}~{\rm
GeV}) \gtrsim 0.3$, that is $g_{a\gamma \gamma} \gtrsim 3 \times
10^{-12}$~GeV$^{-1}$. In what follows, we will refer to this mechanism as
the ``intergalactic conversion'' and use the parameter
constraints~\cite{Galanti-spindex} for this scenario.

The second approach assumes that there are quite strong magnetic fields
inside or around the source, as well as around the observer, while for the
most part of the distance the beam travels in weak magnetic fields,
insufficient for ALP/photon mixing. The conversion may happen either in
the blazar itself and in the Milky Way~\cite{Serpico} or in the galaxy
cluster or filament~\cite{FRT} (see also a more detailed subsequent study
in Ref.~\cite{1207.0776clusters}) containing the source and the observer,
in various combinations.
To get a qualitative idea of the effect of this mechanism on the
gamma-ray attenuation, one might consider the case of maximal mixing,
which is certainly an unrealistic oversimplification for many particular
sources. Then 1/3
of the original photon flux is converted into ALPs close to the source
while the remaining 2/3 attenuate in a usual way. Close to the observer,
2/3 of the ALPs convert back to gamma rays and may be detected. A more
detailed recent study of this mechanism is presented in
Ref.~\cite{Meyer-evidence}, where it is called ``the general-source''
scenario. Notably, this scenario requires $g_{a\gamma \gamma} \gtrsim 2
\times 10^{-11}$~GeV$^{-1}$ because for lower values of the coupling, the
path of the ALP-photon beam would be too short for efficient conversion
even for maximal mixing (the Hillas--like argument). In the rest of the
paper, we refer to this mechanism as the ``galactic conversion'' and use
parameter constraints derived in Ref.~\cite{Meyer-evidence} for this
case\footnote{In Fig.~\ref{fig:main} below, the corresponding line was
smoothed with respect to that of Fig.~4 of Ref.~\cite{Meyer-evidence}
since we believe the features presented there are overprecise, given
observational uncertainty in GMF.}.

Clearly, if we knew the values of IGMF, it would be quite easy to
choose between the two scenarios. Unfortunately, the values are a
subject of strong debates, and we have to seek other ways to disentangle
the two options.

Parameters of the ALP, that is $m$ and
$g_{a\gamma \gamma}$, required for efficient operation of one or another mechanism,
have been discussed in Refs.~\cite{Galanti-spindex, FRT, Meyer-evidence};
we refer the reader to these works for details of the derivation.
Theoretically, they overlap in a large range of allowed parameters.
However, when the most recent experimental and astrophysical constraints
are taken into account, the parameter regions allowed for the two
scenarios become disconnected; this means that if we determine that one or
another scenario works in Nature, we strongly constrain the ALP mass and
coupling! We illustrate this fact in Fig.~\ref{fig:main},
\begin{figure}
\centerline{\includegraphics[width=0.95\columnwidth]{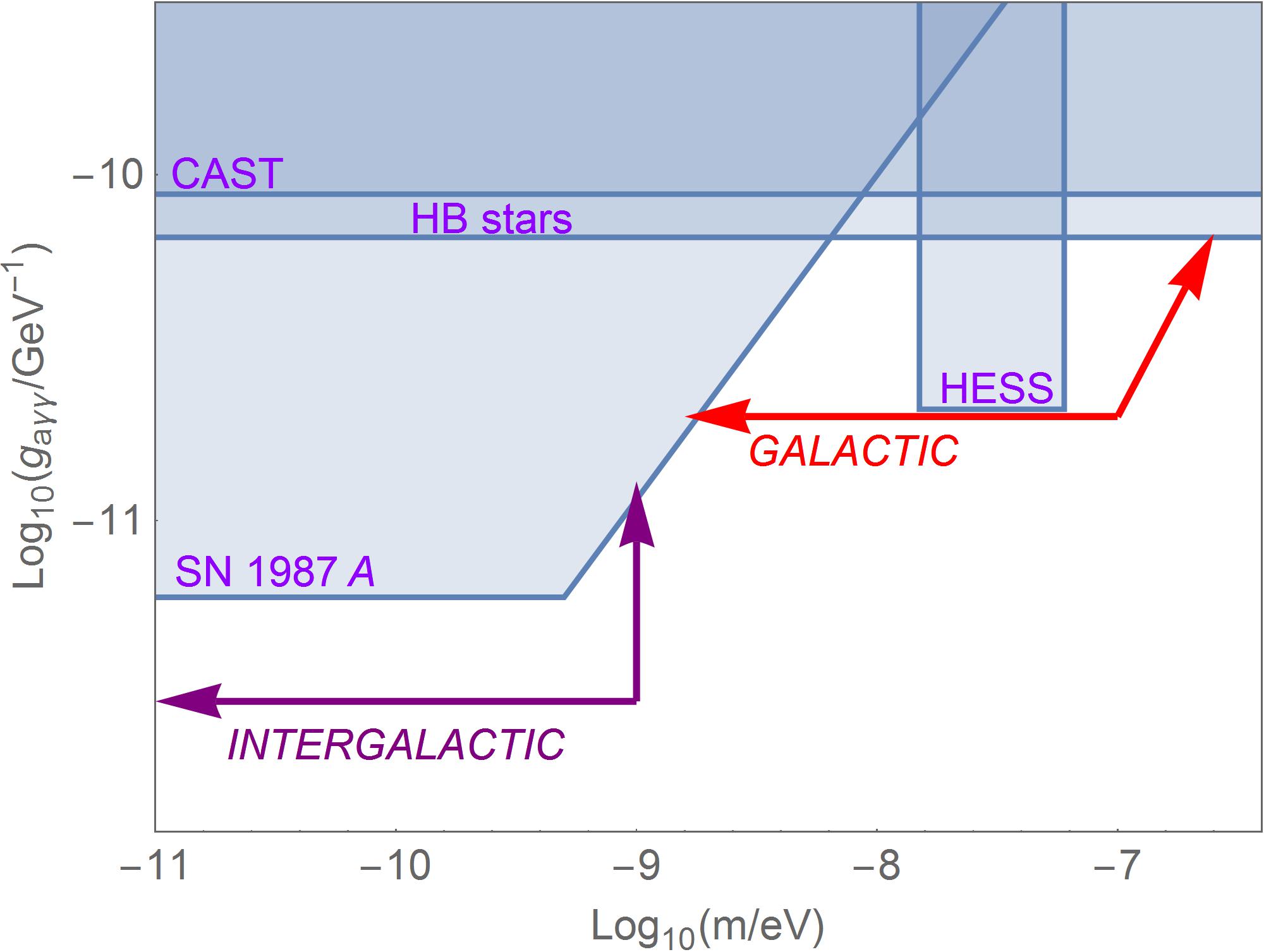}}
\caption{\label{fig:main}
The ALP parameter space (ALP-photon coupling $g_{a\gamma
\gamma}$ versus ALP mass $m$) with current constraints (see text). Regions
corresponding to the Galactic~\cite{Galanti-spindex} and
intergalactic~\cite{Meyer-evidence} ALP/photon conversion explanations of
the gamma-ray anomalies are indicated; they extend to the excluded
regions as shown by arrows. Given all constraints, the two regions are
well separated.}
\end{figure}%
where shaded blue areas, excluded by constraints from the CERN axion
solar telescope (CAST, Ref.~\cite{CAST}), evolution of the
horizonthal-branch (HB) stars~\cite{HBstars}, reanalysis of the
SN~1987A data \cite{SN1987Anew} and HESS constraints from the absence of
irregularities in a blazar spectrum \cite{HESSirreg}, indicate the most
restrictive relevant limits.

One may see that the key constraint contributing to the
separation of the two regions is that of Ref.~\cite{SN1987Anew}. It was
obtained from non-observation of the gamma-ray flare corresponding to the
SN~1987A explosion by the Gamma-Ray Spectrometer of the Solar
Maximum Mission. The fluence limits used for the constraint are 99\% CL;
however, the responce of the instrument to photons at large angles to
viewing directions, relevant for SN~1987A, is quite uncertain. The authors
of Ref.~\cite{SN1987Anew} use results of the previously published
Monte-Carlo simulations. One may ask then, maybe the limit of
Ref.~\cite{SN1987Anew} is wrong and the point of the present paper is
destroyed? The answer is no. Indeed, for $m \lesssim 10^{-9}$~eV, where the
intergalactic scenario may work with $B$ satisfying the observational
limits, the constraint~\cite{SN1987Anew} reads as
$g_{a\gamma \gamma} \lesssim  5 \times 10^{-12}$~GeV$^{-1}$, while the
galactic scenario requires $g_{a\gamma \gamma} \gtrsim 2 \times
10^{-11}$~GeV$^{-1}$, that is four times higher. Since the $g_{a\gamma
\gamma}$ limit changes as the fourth root of the fluence, the two
inequalities for $g_{a\gamma \gamma}$ might be brought into agreement by a
$\sim 4^{4}=256$ times error in the effective area. Clearly, even a rough
estimate of the effective area of a satellite experiment can hardly be a
factor of 256 wrong.

The separation of the two regions, which are often unified in a single
large band referred to as the ``transparency hint'' in relevant plots, is
remarkable. It is instructive to compare this result with those of
Ref.~\cite{Meyer-evidence} which, if considered superficially, suggest
that the two mechanisms may work at the same time. However (see Fig.~4 of
Ref.~\cite{Meyer-evidence}), this might happen only in the ``optimistic''
scenario with the IGMF strength of 5~nG, now excluded by the study of
Faraday rotations of distant sources~\cite{PshirkovIGMF}. We use
parameters from the most recent study of the intergalactic
scenario~\cite{Galanti-spindex} which, though published before
Ref.~\cite{PshirkovIGMF}, does not assume IGMF in excess of the new limit,
1.2~nG. Hence, the probability of conversion in IGMF is lower and the
parameter region is shrinked, compared to Ref.~\cite{Meyer-evidence}.

There exist some concerns about the possibility of efficient axion-photon
conversion in blazars, see Ref.~\cite{Galanti-blazars}. The key
ingredient in this reasoning is the Quantum Electrodynamics (QED)
strong-field term which becomes important at
\begin{equation}
\left(\frac{B}{\rm G} \right)
\left(\frac{E}{\rm 100~GeV} \right)
\sim 0.75,
\label{Eq:QEDinv}
\end{equation}
see e.g.\ Ref.~\cite{FRT} (the ALP-photon mixing in this regime may also
be affected by photon-photon dispersion, see Ref.~\cite{1412.4777}). The
concerns are however not critical for the galactic conversion scenario for
the following reasons. Ref.~\cite{Galanti-blazars} considers separately BL
Lac type objects (BLLs) and flat-spectrum radio quasars (FSRQs). For BLLs,
assuming magnetic fields $B\sim(0.1-1)$~G at the site where gamma rays are
produced, the condition~(\ref{Eq:QEDinv}) is approached at the energies of
interest, so the photon-ALP conversion probability may depend strongly on
unknown details of the environment and becomes hardly predictable.
However, if ALP parameters allow for the conversion in the Milky Way, one
expects that the conversion in a BLL host galaxy or cluster is also
allowed\footnote{One should note, however, that the host galaxies of
blazars are elliptical while the Milky Way is spiral. While the
strength of the magnetic field in the Milky Way is assumed to be
typical, elliptical galaxies contain fewer cosmic-ray electrons, and
hence less synchrotron emission, making detection of the galaxy-scale
magnetic fields tricky and, for today, uncertain \cite{1302.5663}.
Qualitative theoretical arguments suggest \cite{Moss} small correlation
length for the turbulent field. In any case, giant elliptical galaxies
often reside in groups or clusters whose magnetic field also allows for
the photon-ALP conversion.}, and the galactic scenario may work. The
situation is more contrived for FSRQs, for which, in
Ref.~\cite{Galanti-blazars}, magnetic field values $B\sim(1-10)$~G are
assumed (for the gamma-ray emitting region). The QED term starts to
suppress the photon/ALP conversion at energies $E \gtrsim 10$~GeV in this
region, but efficient conversion is possible in outer parts of the galaxy
and the cluster. It may be tricky for gamma rays above $E\sim 20$~GeV to
get there because of intense pair production in the broad-line region,
presumably located between the gamma-ray emitting site and the lobe.
However, one may point out that gamma rays up to several hundred GeV have
been observed from FSRQs~\cite{3c279, PKS1222+21} (see
Ref.~\cite{PKS1222+21+} for a discussion of ALP explanation of the
observation~\cite{PKS1222+21}), which means they escape the broad-line
region somehow. Once escaped, thay may equally well convert to ALPs
outside.

\section{Discrimination
between galactic and intergalactic scenarios}
\label{sec:discri}
{\bf Anisotropy.} The magnetic field of the Milky Way galaxy has a
complicated structure, and the probability of the ALP/photon conversion
there, which is required in the galactic scenario, depends strongly on the
direction. Evidence for direction dependence in the anomalous transparency
of the Universe may therefore be a strong argument in favour of the
galactic scenario~\cite{Serpico, FRT, Brun-aniso}.

In Ref.~\cite{Serpico}, it was pointed out that the positions of a few TeV
blazars with redshifts $z>0.1$ known by that time fit surprisingly well the
regions in the sky where the conversion probability, calculated within the
model of the Galactic magnetic field (GMF) of Ref.~\cite{GMF}, are high.
Here, we assume this as a hypothesis and attempt to test it with new
observational data. Clearly, more elaborated approaches should be used in
further studies (see Ref.~\cite{Brun-aniso} for an attempt on which we
comment below and Ref.~\cite{in-progress} for a different approach). We
consider a sample of blazars with firm detection beyond $\tau=1$ which
consists of 15 objects observed by IACTs and 5 objects observed by FERMI
LAT (the sample of Ref.~\cite{gamma}), supplemented by additional 6
blazars rejected in Ref.~\cite{gamma} because of the insufficient number
(4 with 5 required) of data points for fitting spectra with breaks. We
drop 4 nearby objects with $z<0.1$ from the sample, like it was done in
Ref.~\cite{Serpico}.

\begin{figure}
\centerline{\includegraphics[width=0.9\columnwidth]{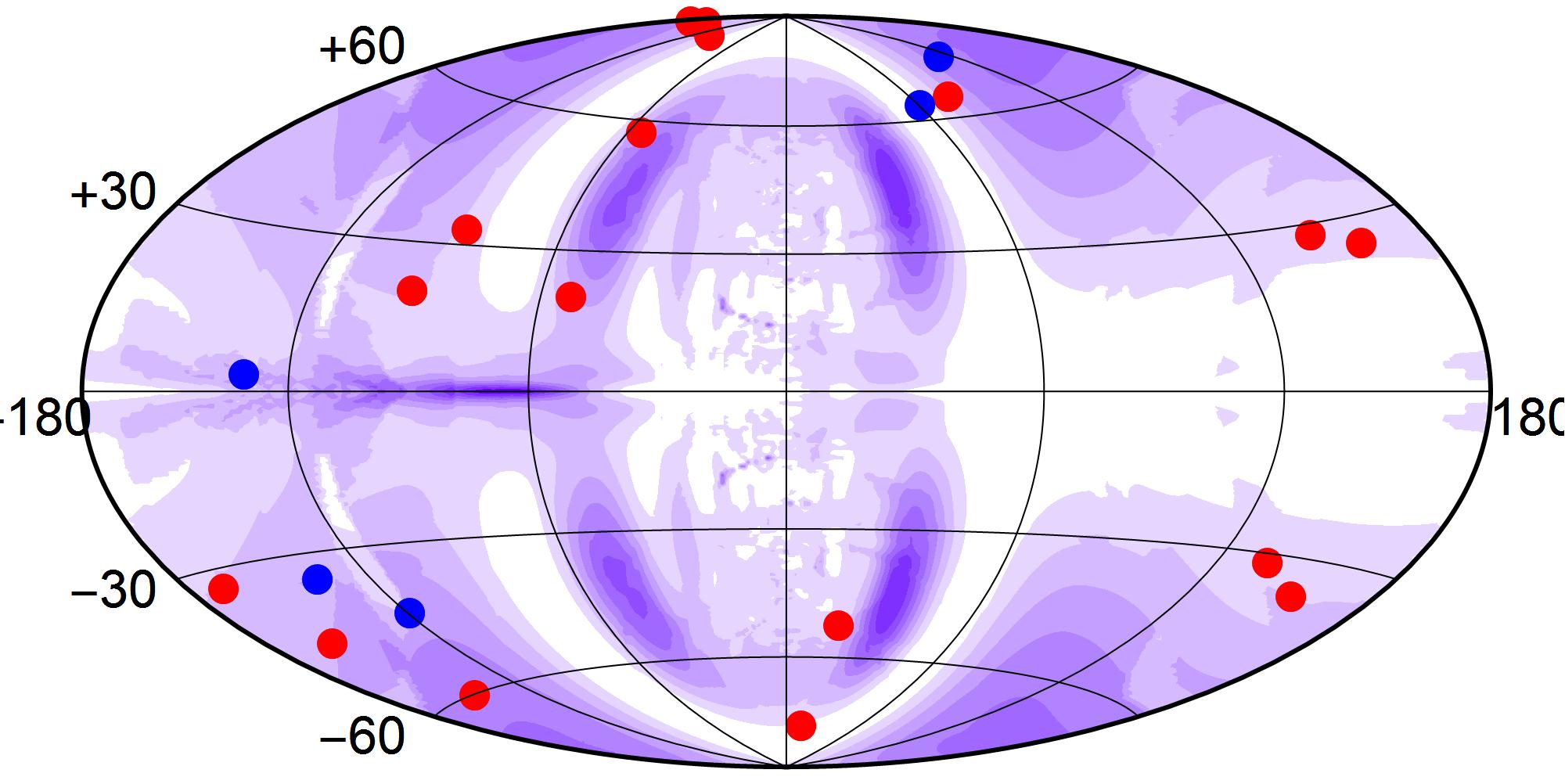}
~~~
\includegraphics[width=0.1\columnwidth]{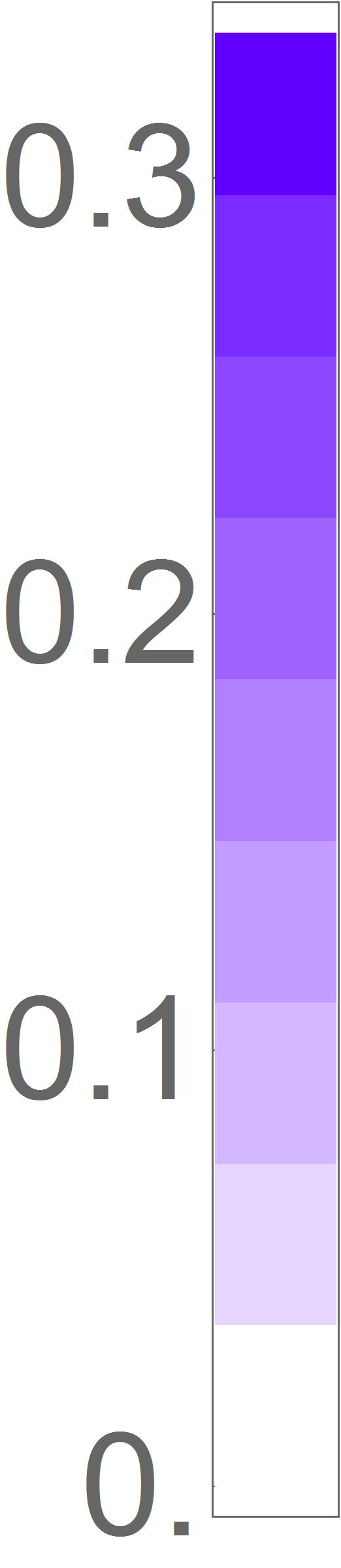}}
\caption{\label{fig:map}
The skymap (Galactic coordinates, Hammer projection) with positions of
blazars with detected gamma-ray flux at energies for which $\tau>1$ (red,
$0.1<z<1$; blue, $z>1$), see text. Deeper shading corresponds to
higher ALP-photon conversion probability in the Milky Way (the Galactic
magnetic field model of Ref.~\cite{GMF}).}
\end{figure}

Figure~\ref{fig:map}
represents the distribution of these objects in the sky together with the
conversion probability for the same magnetic-field model~\cite{GMF}.
Though there exist more elaborated modern GMF models
\cite{PshirkovGMF, FarrarGMF}, the test of the original claim should be
performed with the same one. The objects indeed follow the regions of high
conversion probability, qualitatively confirming the trend seen in
Ref.~\cite{Serpico}. Two of 22 objects are seen in the regions of low
conversion probability; this may be well understood given uncertainties in
the GMF models.

\begin{figure}
\centerline{\includegraphics[width=0.95\columnwidth]{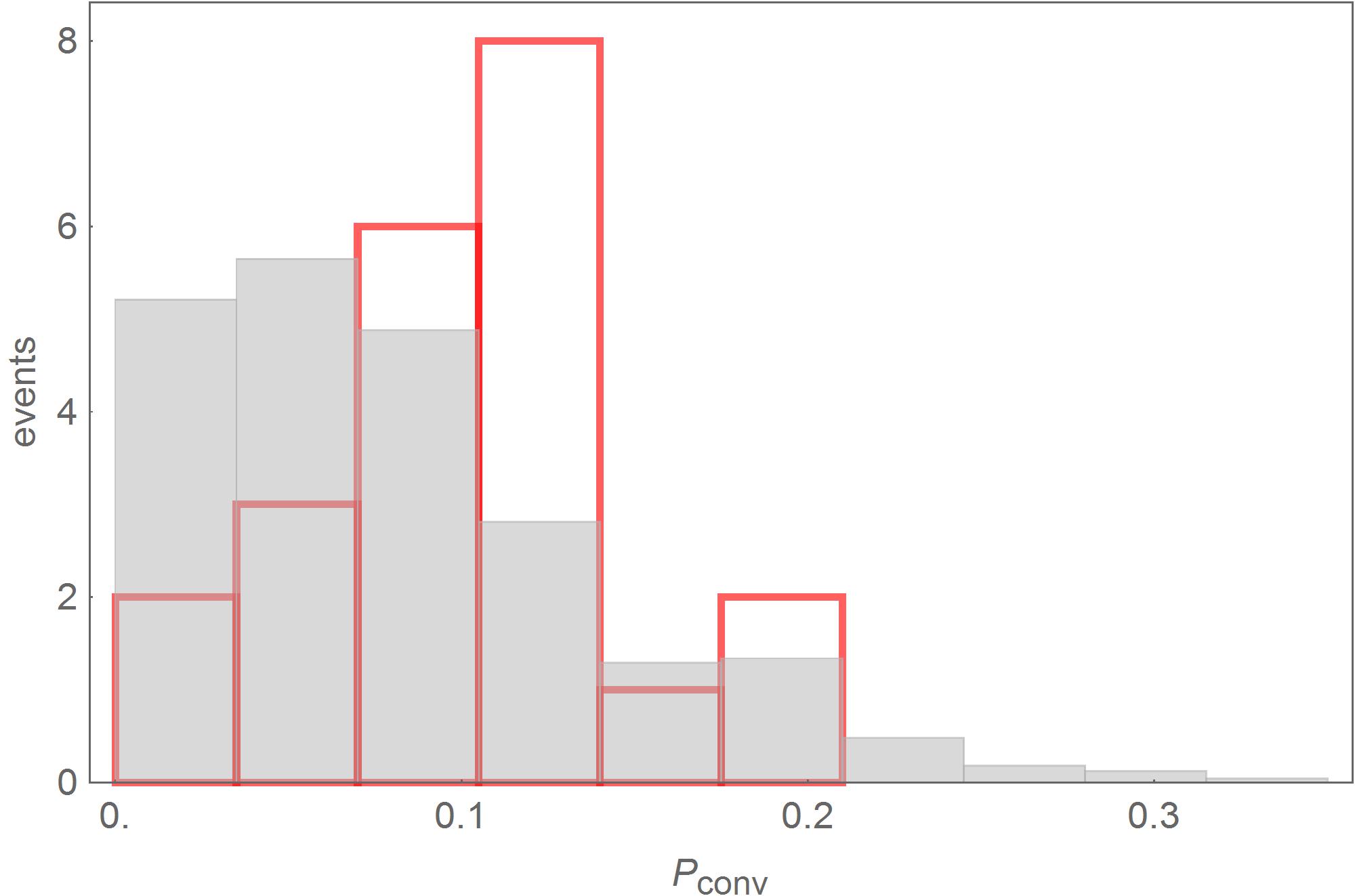}}
\caption{\label{fig:histo}
The distribution of the values of the ALP-photon conversion probability in
the Milky Way calculated for directions to blazars with detected gamma-ray
flux at energies for which $\tau>1$ (red line) and for random isotropic
directions (gray shading). }
\end{figure}

It is not possible, however, to rigorously test the hypothesis
quantitatively, because the blazars we discuss do not form a complete
isotropic sample. While 5 FERMI-LAT sources were selected from a more or
less uniform full-sky data, the remaining 17 objects were arbitrarily
chosen for observations with pointed runs of small-field-of-view IACTs.
Nevertheless, for illustration, we present here the results of a simple
statistical test of the hypothesis, keeping in mind its qualitative level.
For each of the 22 sources in the sample, we calculate the ALP/photon
conversion probability in the GMF of Ref.~\cite{GMF}.
The same distributions were calculated and averaged for 100 sets of 22
objects distributed isotropically in the sky. Figure~\ref{fig:histo}
compares the distributions. The Kolmogorov-Smirnov probability that the
distribution seen for the real data is a fluctuation of that for simulated
directions is 0.02. We do not assign any statistical significance to this
result for the reasons mentioned above (incomplete sample), though the
entire picture does not contradict the galactic conversion scenario.

Another approach to test the anisotropy effect was suggested by Wouters
and Brun~\cite{Brun-aniso}. They proposed to study autocorrelation
patterns in the directional distribution of differences between FERMI-LAT
and TeV spectral indices of blazars. We note that these differences are
not equal to the spectral breaks discussed in Sec.~\ref{sec:evidence}
since the breaks happen at distance-dependent energies, not always between
the bands considered in Ref.~\cite{Brun-aniso}. The procedure is motivated
by the fact that, due to our limited knowledge of the GMF, actual
conversion probabilities vary strongly from one GMF model to another,
while all models predict patchy large-scale anisotropy in the distribution
of these probabilities. However, as they show
themselves~\cite{Brun-aniso}, the patterns in the autocorrelation function
also vary strongly from model to model. Clearly, quantitative estimates
of statistical significance would require a complete sample of sources in
this approach, similarly to a more direct approach~\cite{Serpico} we
discuss here.

{\bf Distant objects.} In the ideal case and in the long-distance limit,
the effective optical depth $\tau_{\rm ALP}$ behaves differently in the
two scenarios: for intergalactic conversion, $\tau_{\rm ALP} \sim
(2/3)\tau$ (and therefore grows approximately linearly with distance,
like the standard optical depth $\tau$), while for the galactic scenario,
assuming maximal mixing, it reaches a constant, distance-independent value
corresponding to the flux suppression by a factor of $\sim 2/9$, that is
$\tau_{\rm ALP} \sim 1.5$. At a certain redshift $z_{\rm crit}$, which
value depends on the details of the absorption model and of magnetic
fields assumed, the two suppression factors are equal, while beyond
$z_{\rm crit}$, the effective absorption becomes stronger and stronger in
the intergalactic scenario, remaining essentially constant in the galactic
one. This means that for very high redshifts, the anomalous transparency
effect would hardly be seen in observations for the intergalactic
scenario, therefore any evidence for the effect for very distant sources
speaks in favour of the galactic conversion. The blazars observed by IACTs
have measured spectroscopic redshifts up to $z=0.536$ (3C~279; a lower
limit of $z>0.6$ exists for PKS~1424$+$240), and all
anomalous-transparency effects derived from them are equally well
described by both mechanisms. However, inclusion of much more distant
blazars, for which the absorption becomes significant at energies in the
FERMI-LAT band, changes the picture: the 12-sigma anomaly
reduces to $\sim 5\sigma$ (and therefore remains present) when
intergalactic conversion is assumed, in addition to the usual absorption,
but diminishes to $\sim 2\sigma$ (and therefore disappears) for the
assumption of the galactic scenario~\cite{gamma}. While these results have
been obtained in a simplified approach, the difference between scenarios
is so pronounced that it could hardly be removed by any detalisation of
the analyzis. Prospects for observations of distant gamma-ray sources are
discussed, in the ALP context, in Ref.~\cite{1410.1556}.

{\bf Intergalactic magnetic fields.}  The intergalactic scenario requires
rather high IGMF, $B\sim(10^{-10} - 10^{-9})$~G, otherwise the conversion
probability would be too low.
The value of IGMF is irrelevant for the galactic scenario provided the
SN~1987A constraints on ALP parameters are satisfied. Present-day knowledge
does not allow for a firm conclusion about the real values of $B$. A
number of constraints are summarized in the
review~\cite{Neronov-IGMF-rev}. The most stringent observational limit,
based on the redshift independence of the Faraday rotation from distant
sources, is $B\le 1.2 \times 10^{-9}$~G \cite{PshirkovIGMF}. Constrained
simulations of IGMF \cite{Dolag} favour the values of $B\sim (10^{-12} -
10^{-11})$~G in voids (see however Refs.~\cite{0102076, Sigl, 0604462},
advocating somewhat larger values from unconstrained simulations and
semi-analitical models). The angular correlation function of FERMI-LAT
photons points to $B\sim 5 \times 10^{-14}$~G \cite{Tashiro2014}. There
exist several claims of observations of the pair halo around gamma-ray
sources which suggest IGMF values in the range of $B\sim 10^{-14}$~G, see
e.g.\ Ref.~\cite{Chen2014} for a recent one, but these analyses are
technically involved and require further confirmation.

\section{Future tests}
\label{sec:future}
While all three methods to distinguish between the two scenarios,
discussed in Sec.~\ref{sec:discri}, favour weakly the galactic conversion
mechanism, it is clear that more tests are required both to confirm the
anomalous transparency of the Universe and to single out its explanation.
To approach the tests on more solid grounds, future observations are
necessary.
Of particular importance are spectral and anisotropy studies, for which
the following directions are especially important:
\begin{itemize}
 \item
to enlarge the overall statistics of TeV blazars, which is best achieved
with the coming Cerenkov Telescope Array (CTA) \cite{CTA};
\item
to study absorption effects in the spectra of the most distant blazars,
for which one needs high-sensitivity observations at energies $\sim
(10-100)$~GeV. The sensitivities of both FERMI LAT and CTA are
insufficient in this energy range; the solution may be provided by
high-altitude low-threshold Cerenkov detectors \cite{low-threshold}.
Presently, two projects of this kind are under consideration, the
Atmospheric Low Energy Gamma-Ray Observatory (ALEGRO) in Atacama, Chile,
and the Elbrus Gamma-Ray Observatory (EGO) at the Mount Elbrus, Russia;
\item
to move into the strong-absorption energy range for bright nearby blazars,
which would require observations at $\sim 100$~TeV. The proper instruments
for that would be extensive-air-shower detectors, in particular, the Tunka
Advanced Instrument for cosmic-ray physics and Gamma-ray Astronomy (TAIGA)
\cite{TAIGA} and the upgraded Carpet array at the Baksan Neutrino
Observatory \cite{Carpet3} in the nearest future. Several years later, the
Large High Altitude Air Shower Observatory (LHAASO) will provide the best
sensitivity \cite{LHAASO}. Another relevant future project is HiSCORE
\cite{1108.5880}.
\end{itemize}

Additional important contributions to the discussion are expected from
observational constraints on the IGMF values. Note that the improvement in
the limits on $B$ by a factor of two would be sufficient to
independently exclude the intergalactic scenario. Such an improvement
may be achieved after the data of the planned all-sky rotation-measure
grid~\cite{Beck:2004gq, Beck:2008qf}, planned for the Square Kilometer
Array (SKA), will become available. Of course, constraints on $g_{a \gamma
\gamma }$ from laboratory searches for the responsible ALP would be
crucial, with the most sensitive planned instrument being the International
Axion Observatory (IAXO) \cite{IAXO}, for which the entire
range of photon-ALP couplings suggested in both scenarios we discussed is
within the discovery range. The upgraded Any Light Particle Search
(ALPS-II) \cite{ALPS3} experiment will approach the interesting range
of couplings and probe a part of the parameter space for the galactic
scenario.

\begin{acknowledgements}
The author is indebted to G.~Galanti, M.~Meyer,
M.~Pshirkov, M.~Roncadelli and G.~Rubtsov for interesting discussions.
This work was supported in part by the RFBR grant 13-02-01293.
\end{acknowledgements}

\end{document}